\documentclass[12pt]{article}
\usepackage{geometry}                
\usepackage{graphicx}
\usepackage{amssymb}
\usepackage{epstopdf}
\usepackage{amsmath}
\usepackage{cite}
  
\input xy
\xyoption{all}

\def\toclevel@subsubsubsection{4}
\def\toclevel@paragraph{5}
\def\toclevel@paragraph{6}
\def\l@subsubsubsection{\@dottedtocline{4}{7em}{4em}}
\def\l@paragraph{\@dottedtocline{5}{10em}{5em}}
\def\l@subparagraph{\@dottedtocline{6}{14em}{6em}}

\def\beq{\begin{equation}}
\def\eeq{\end{equation}}
\def\beqn{\begin{eqnarray}}
\def\eeqn{\end{eqnarray}}

\def\T{{\rm T}}

\newcommand{\sma}{\left(\begin{smallmatrix}}
\newcommand{\smaa}{\end{smallmatrix}\right)}

\newcommand{\gsim}{\lower.7ex\hbox{$
\;\stackrel{\textstyle>}{\sim}\;$}}
\newcommand{\lsim}{\lower.7ex\hbox{$
\;\stackrel{\textstyle<}{\sim}\;$}}

\begin{document}

\begin{titlepage}

\begin{flushright}
FTPI-MINN-14/35, \\
UMN-TH-3406/14\\
\end{flushright}

\vspace{1cm}

\begin{center}
{  \Large \bf  On Isometry Anomalies in Minimal \boldmath{${\mathcal N}=(0,1)$} \\[2mm]  and \boldmath{${\mathcal N}=(0,2)$} Sigma Models}

\vspace{0.8cm}

{\large
Jin Chen,$^{a}$ Xiaoyi Cui,$^{b}$ Mikhail Shifman,$^{a,c}$ \\[2mm] and Arkady Vainshtein$^{\,a,c}$}
\end {center}

\vspace{1mm}

\begin{center}

$^{a}${\it  Department of Physics, University of Minnesota,
Minneapolis, MN 55455, USA}\\[1mm]
$^{b}${\it Mathematisches Institut, Georg-August Universit\"{a}t G\"{o}ttingen, G\"{o}ttingen, D-37073, Germany}\\[1mm]
$^c${\it  William I. Fine Theoretical Physics Institute,
University of Minnesota,
Minneapolis, MN 55455, USA}\\[1mm]

\end{center}

\vspace{0.5cm}

\begin{center}
{\large\bf Abstract}
\end{center}

The two-dimensional minimal supersymmetric sigma models with homogeneous target spaces $G/H$ and chiral fermions of the same chirality are revisited. We demonstrate that the Moore-Nelson consistency
condition revealing a global anomaly in ${\rm CP}(N\!-\!1)$  (with $N\!>\!2$ and ${\mathcal N}\!=\!(0,2)$ supersymmetry) due to a nontrivial first Pontryagin class is in one-to-one correspondence with the local anomalies of isometries in these models. These latter anomalies are generated by fermion loop diagrams which we explicitly calculate.
In the case of ${\rm O}(N)$ sigma models the first Pontryagin class vanishes, so there is no global obstruction
for the minimal ${\mathcal N}=(0,1)$ supersymmetrization of these models. We show that at the local level
isometries in these models are anomaly free. Thus, there are no obstructions to quantizing the
  minimal ${\mathcal N}=(0,1)$ models with the ${\rm S}^{N\!-\!1}\!= {\rm SO}(N)/{\rm SO}(N\!-\!1)$ target
  space. This also includes ${\rm CP}(1)$ (equivalent to ${\rm S}^{2}$) which is  an exceptional case from the ${\rm CP}(N\!-\!1)$ series. We also discuss a relation
between the geometric and gauged formulations of the ${\rm CP}(N\!-\!1)$ models.

\hspace{0.3cm}

\end{titlepage}


\newpage


\section{Introduction}
\label{OnCPn}

Two-dimensional chiral sigma models are known for a long time (e.g. \cite{1}). Recently chiral $\mathcal{N}\!=(0,2)$ sigma models
emerged as low-energy world sheet theories on non-Abelian strings supported in some $\mathcal{N}\!\!=\!1$
four-dimensional Yang-Mills theories \cite{Edalati:2007vk, Shifman:2008wv} (for a review see  \cite{Shifman:2014jba}). This explains a renewed interest in their studies. Recent works include (but are not limited to) Refs.
 \cite{Witten:2005px,bai2,bai3,Adams:2003zy,Melnikov:2012hk,Jia,Gadde:2013lxa,Gadde:2014ppa,Cui:2011rz,
 CCSV1,Cui:2010si,Cui:2011uw,Shifman:2008kj,B1}.
 Most of these papers are devoted to nonminimal chiral models.
 Nonminimal chiral models are obtained as deformations of $\,\mathcal{N}\!=\!(2, 2)$
 supersymmetries and contain both \emph{left} and \emph{right}-handed fermions.\footnote{\,Strictly speaking, in two dimensions
  they are {\it left}- and {\it right}-movers.}
  They are
 free of anomaly by construction. In this paper, we consider {\em minimal }chiral sigma models
 with $\mathcal{N}\!=(0, 1)$ and $(0, 2)$ supersymmetries  where by ``minimal'' we mean that there are only,
 say,  \emph{left}-handed fermions included in models. Since these minimal chiral theories generically suffer
 from anomalies, the no-anomaly conditions become criteria for mathematical consistency of these models {\em per se}.
 In general the (\emph{left}-handed) chiral fermions can be defined on arbitrary vector bundles over manifolds on which
 bosonic fields live in various dimensions. There are two types of intrinsic anomalies in the minimal ${\mathcal N}=(0,2)$
 sigma models, which can be compared to those in gauge theories in
 four dimensions.

First,  chiral fermions in four dimensions ($4d$) can ruin gauge invariance already at one loop, as it happens, say,  in the ${\rm SU}(N)$ gauge theory with $N>2$ and a single chiral fermion in the fundamental representation. This anomaly does not appear, however, in the ${\rm SU}(2)$ gauge theory due to the absence of the $d$ symbols in ${\rm SU}(2)$. Nevertheless, the ${\rm SU}(2)$ gauge theory with one chiral fermion in the fundamental representation does not exist since it suffers from a ``global" Witten's anomaly \cite{su2ga}. This is the second type of anomalies in four-dimensional Yang-Mills.

The anomalies in the minimal $\mathcal{N}\!=(0,2)$ sigma models  were discussed by many authors in different aspects. A number of authors calculated \cite{Zumino:1983rz, Bagger:1985, Manohar:1984uq, Manohar:1984zj, AlvarezGaume:1985yb} chiral anomalies (i.e. obtained explicit local forms) and discussed the mechanism of the anomaly cancellations. On the other hand, global feature of anomalies were also thoroughly considered  in the works \cite{Moore:1984ws, AlvarezGaume:1984dr}. A well-known no-go theorem
 \cite{Moore:1984ws} establishing a global anomaly due to non-zero Pontryagin classes over vector bundles in
such minimal $(0, 2)$ models such as ${\rm CP}(N\!-\!1)$  makes them inconsistent (with an exception of ${\rm CP}(1)$ model). We will revisit this issue. In the present paper we will discuss
 the isometries of the target space manifolds ${\rm O}(N)$ and ${\rm CP}(N\!-\!1)$ and the corresponding isometry anomalies. In the subsequent publication we will discuss more general aspects, such as anomalies vs.
diffeomorphism/holonomy invariance of the target space after quantization in generic sigma models.

A few words on terminology. Sigma models are defined on manifolds which typically have to be covered by many local patches.
One can specify a local chart of the manifold and then perform the anomaly calculation (typically at one loop).
We thus call such anomalies \emph{local}.\footnote{
\,The isometries in the sigma models under consideration are global symmetries analogous to
flavor symmetries in the gauge theories. We will still refer to the isometry anomalies as to local anomalies, to avoid confusion.}

In the case of local anomalies offset by
counterterms on local patches, one must worry how to patch these counterterms on different charts.
It is essentially a cohomology problem  \cite{Alvarez:1985} which thus is tied up with  global features of the manifolds under consideration.
A large class of sigma models admits the so-called gauged formulation  \cite{Witten:1979} (e.g. variants of the  Grassmannian
sigma models).   In the gauged formulation we potentially have to deal with chiral anomalies in the ``small" and  ``large'' gauge transformations
(analogous to the Witten ${\rm SU}(2)$ anomaly \cite{su2ga}).

The target spaces in the problems to be considered are homogeneous symmetric spaces of $G/H$ type.
In this case it was shown \cite{Manohar:1984zj,Moore:1984ws} that the criteruim of local anomalies is stronger than
the global obstruction: the local anomalies imply the global obstruction and {\em vice versa}. In our examples of
the  ${\rm O}(N)$ and ${\rm CP}(N\!-\!1)$ models we explicitly verify this statement.
In the first  example which is free from the Moore-Nelson global obstruction,  we demonstrate
that the  ${\rm O}(N)$ model is free from local anomalies. In the second example, ${\cal N} = (0,2)$ ${\rm CP}(N\!-\!1)$ models, $N>2$, in which the first Pontryagin class is nontrivial, we found local anomalies.
 Thus, such models are inconsistent.

As was mentioned, in both cases we examine
anomalies in the isometries which decide whether or not geometry of the
classical action can be maintained at the quantum level. Only if it can be maintained can the theory be self-consistent.
In the minimal ${\rm O}(N)$ models one can construct anomaly-free isometry currents,
while such a construction is impossible in the minimal ${\rm CP}(N\!-\!1)$ models. The only exception is ${\rm CP}(1)$ which is equivalent to ${\rm O}(3)$.

The paper is organized as follows.
In Sec.\ \ref {On} we thoroughly discuss the minimal ${\rm O}(N)$ models and demonstrate the absence of the isometry anomalies.  Section \ref{cpsm} is devoted to the minimal ${\rm CP}(N\!-\!1)$ models. We derive
the ${\rm CP}(N\!-\!1)$  isometry anomalies in this model. Our analysis in this section is somewhat different from the ${\rm O}(N)$ case. We examine the correspondence between the isometry anomalies in non-linear sigma models (NLSM) and gauge anomalies in gauged linear sigma models (GLSM), and then derive the isometry anomalies based on the above correspondence. In Sec.\ \ref{dual} we consider a dual formalism for the ${\rm O}(N)$ models
and arrive at the same result as in Sec.\ \ref {On} by using the correspondence referred in this section.
Section \ref{concl} summarizes our results and outlines questions for future explorations. Appendix presents details of derivation in Sec.\ \ref{cpsm} through direct calculation as a verification.

\section{\boldmath{${\rm O}(N)$} Sigma Model}
\label{On}

Let us first study the ``linear" version\footnote{By linear we mean that the kinetic term of Lagrangian in ${\rm O}(N)$ model is linear. The model is, for sure, subject to the constraint $n^in_i=1$ which makes it nonlinearly realized when we remove the extra redundancy by solving the constraint.} of the ${\rm O}(N)$ model \cite{Novikov:1984}, investigate the symmetries of the model and then pass to the nonlinear description.

The linear ${\rm O}(N)$ sigma model contains $N$ real fields $n^i$, where $i=1,2,...,N$, with the constraint
\beq
 n^{i}n_i=1\,.
 \label{c1}
 \eeq
This means that the target space of the model is the sphere ${\rm S}^{N\!-\!1}$,
which could be viewed as the coset
 \beq
{\rm S}^{N-1}={\rm SO}(N)/{\rm SO}(N\!-\!1)\,.
\label{nover}
 \eeq
 Thus, the model (\ref{lb}) can equally be referred to as the ${\rm S}^{N\!-\!1}$ model. In the literature the first name, ${\rm O}(N)$, is more common, however. It reflects, in particular, that counting of isometries is given by ${\rm O}(N)$.
The bosonic part of Lagrangian is
\begin{eqnarray}
\mathcal
{L}_{b}=\frac{1}{2g^2_0}\,\partial_{\mu}n^i\partial^{\mu}n_i+\lambda(n^i n_i-1)
\label{lb}
\end{eqnarray}
where $\lambda$ is a Lagrange multiplier that ensures the  constraint
above on the  $n^i$ fields.
As mentioned above there are   $N(N-1)/2$ isometry symmetries
corresponding to the ${\rm SO}(N)$ group. For each point in the target
space a stationary subgroup $H={\rm SO}(N\!-\!1)$ (the denominator in
Eq.\,(\ref{nover})) consists of transformations which do not act at
this point. We fix a particular choice of $H$ specifying an axis, say
$n^N$, as associated with the stationary under ${\rm SO}(N\!-\!1)$ point. For other points the
transformations from $H$ are realized linearly.

Thus, the first set of isometries is given by linear transformations,
\begin{equation}
\delta_{\epsilon}n^i=\epsilon^{ij}n_j, \ \ \ \delta_{\epsilon}n^N=0,
\end{equation}
where $\epsilon^{ij}=-\epsilon^{ji},\ (i=1,2,...,N\!-\!1)$ are infinitesimal parameters.
The remaining $N\!-\!1$ isometries form the second set where the
transformations are realized nonlinearly,
\begin{equation}
\delta_{\alpha}n^i=\alpha^{i}n^N, \ \ \ \delta_{\alpha}n^N=-\alpha^{i}n_i\,,
\end{equation}
where $\alpha^i,\ (i=1,2,...,N\!-\!1)$ are infinitesimal parameters. The sub/superscripts  are raised or lowered by $\delta^{ij}$ or $\delta_{ij}$.

Now, one can rewrite this sigma model through the standard stereographic projection to explicitly solve the constraint $n^i n_i=1$, by setting
\begin{eqnarray}
\phi^i=\frac{n^i}{1+n^N}\,, \ \ \ i=1,2,...,N-1\,.
\label{stereoproj}
\end{eqnarray}
By recalculating the infinitesimal transformations of $\phi^i$ with respect to $\epsilon^{ij}$ and $\alpha^i$, one obtains
\begin{equation}
\begin{split}
&\delta_\epsilon\phi^i=\epsilon^{ij}\phi_j\,,\\[1mm]
&\delta_\alpha\phi^i=\frac{1-\phi^2}{2}\,\alpha^i+\alpha^j\phi_j\phi^i\,.
\end{split}
\label{isotranf}
\end{equation}
In terms of the field $\phi^i$ the Lagrangian (\ref{OnLagrangian}) takes the form,
\begin{equation}
\mathcal{L}_{b}=\frac{1}{2}\,g_{ij}\partial_{\mu}\phi^i\partial^{\mu}\phi^j\,,
\end{equation}
where $g_{ij}$ is the metric tensor of $S^{N\!-\!1}$ sphere,
\begin{equation}
g_{ij}=\frac{4}{g^2_0}\,\frac{\delta_{ij}}{(1+\phi^2)^2}\,.
\label{metr}
\end{equation}

Supersymmetrizing the  ${\rm O}(N)$ Lagrangian by adding left-handed fermions is straightforward.
We couple $N\!-\!1$ real {left}-handed chiral fermions $\psi^i\equiv\psi^i_L$ to the bosonic fields so that the theory has $\mathcal{N}=(0,1)$ supersymmetry,
\begin{eqnarray}
\mathcal{L}=\mathcal{L}_{b}+\mathcal{L}_{f}=\frac{1}{2}\,g_{ij}\partial_{\mu}\phi^i\partial^{\mu}\phi^j
+\frac{i}{2}\,g_{ij}\bar{\psi^i}\gamma^{\mu}D_{\mu}\psi^j
\label{OnLagrangian}
\end{eqnarray}
where $D_{\mu}$ is the covariant derivative pulled back from the
$S^{N\!-\!1}$ sphere. The stereographic projection (\ref{stereoproj}) gives
us a local chart $\{\phi^i\}$, for which one can write down the metric (see (\ref{metr})) and connections explicitly,
\begin{equation}
\begin{split}
D_\mu\psi^i=\partial_\mu\psi^i+\Gamma^i_{jk}\partial_{\mu}\phi^j\psi^k\,,\,\qquad\qquad\\[1mm]
\Gamma^i_{jk}=-\frac{2}{1+\phi^2}\,(\delta^i_j\phi_k+\delta^i_k\phi_j-\delta_{jk}\phi^i)\,.
\label{roundmetric}
\end{split}
\end{equation}

\vspace{2mm}

Now let us pass to the issue of isometry anomalies.
To evaluate the anomalies, one needs to integrate out fermions to find the effective action $\Gamma_{\rm eff}[\phi]$. Then one performs the isometry transformations (\ref{isotranf}). We introduce  vielbeins $e^a_{\ i}$ on $S^{N-1}$ to decompose the metric and rewrite fermion fields in the canonic way,
\begin{eqnarray}
e^a_{\ i}=\frac{2}{g_0}\,\frac{1}{1+\phi^2}\,\delta^a_{\ i}\,, \qquad e^i_{\ b}=\frac{g_0}{2}\,(1+\phi^2)\,\delta^i_{\ b}\,.
\label{8}
\end{eqnarray}
Apparently $e^a_{\ i}$ satisfy the conditions
\begin{eqnarray}
e^a_{\ i}e^i_{\ b}=\delta^a_{\ b}\,, \qquad \delta_{ab}e^a_{\ i}e^b_{\ j}=g_{ij}\,.
\label{vielbein}
\end{eqnarray}
In Eqs.\,(\ref{8}) and  (\ref{vielbein}) $\delta_{ab}$ and $\delta^{ab}$ are for raising and lowering indices $\{a, b, ...\}$, while $g_{ij}$ and $g^{ij}$ for indices $\{i, j, ...\}$. Besides, for local chart $\{\phi^i\}$, one can still use $\delta_{ij}$ to write $\phi_i\equiv\delta_{ij}\phi^j$.

\vspace{2mm}

As long as conditions (\ref{vielbein}) are met, one still has a residual freedom to make different
choices for $e^a_{\ i}$. This freedom might lead to the so-called holonomy anomalies which we will discuss
in upcoming work \cite{CCSV3}.

\begin{figure}[h]
\begin{center}
\includegraphics[width=6cm]{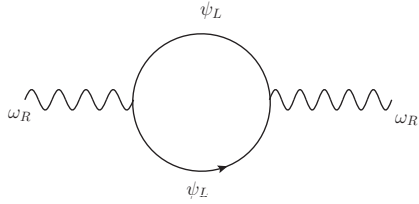}
\end{center}
\caption{\small The wavy lines denote external spin-connection fields $\omega_R$, and solid lines denote chiral fermion $\psi_L$.}
\label{Biangle}
\end{figure}

Through vielbeins $e^a_{\ i}$ we define $\psi^{a}\equiv e^a_{\ i}\psi^i$, and thus rewrite the fermion part of the ${\mathcal N}=(0,1) $ Lagrangian,
\begin{eqnarray}
\mathcal{L}_f=\frac{i}{2}\,g_{ij}\bar{\psi^i}\gamma^{\mu}D_{\mu}\psi^j=\frac{i}{2}\,\bar{\psi^a}\gamma^{\mu}(\partial_{\mu}\delta_{ab}+\omega_{abi}\partial_{\mu}\phi^i)\psi^b\,,
\end{eqnarray}
where
\begin{equation}
\omega^a_{\ bi}=e^a_{\ j}{\cal D}_ie^j_{\ b}=e^a_{\ j}\Bigg[\frac{\partial e^j_{\ b}}{\partial \phi^{i}}+\Gamma^{j}_{ik}e^k_{\ b}\Bigg]
\label{ome}
\end{equation}
is the spin-connection on the frame bundle, and ${\cal D}_{i}$ is the covariant derivative on $S^{N-1}$. Now, we integrate out fermions and arrive at an effective action $\Gamma_{\rm eff}$. This requires calculation
of the  bi-angle diagram (see Fig.\,\ref{Biangle});  higher orders are finite and thus do not contribute into anomalies.
Note that there are only chiral fermions $\psi^a_L$ in $\mathcal{L}_f$ coupled to the spin-connection $\omega_R=\omega_i\partial_R\phi^i$. Therefore, the effective action is a functional of $\omega_R$,
\begin{eqnarray}
i\Gamma_{\rm eff}[\omega_R]&=&\frac{i}{16\pi}\int d^2x\ \omega^a_{\ b\mu}\left(g^{\mu\alpha}+\epsilon^{\mu\alpha}\right)\frac{\partial_\alpha\partial_\beta}{\partial^2}\left(g^{\beta\nu}-\epsilon^{\beta\nu}\right)\omega^b_{\ a\nu}+\mathcal{O}(\omega^3_R)\nonumber\\[1mm]
&=&\frac{i}{16\pi}\int d^2x\ \omega^a_{\ bR}\frac{\partial_L\partial_L}{\partial^2}\,\omega^b_{\ aR}+\mathcal{O}(\omega^3_R)\,.
\label{Leff}
\end{eqnarray}

As we mentioned above cubic and higher term in the action are given by well convergent integrals in momentum space.
It means that these terms are well defined in UV and anomaly could come only from quadratic in connections terms.

To evaluate the isometry anomalies   from $\Gamma_{\rm eff}$, we will examine $\delta_\epsilon\Gamma_{\rm eff}$ and $\delta_\alpha\Gamma_{\rm eff}$ under isometry transformations (\ref{isotranf}).
Invariance of $\Gamma_{\rm eff}$ under linear transformations, $\delta_\epsilon\Gamma_{\rm eff}=0$
is evident because these symmetries are explicitly maintained. As for nonlinear transformations we will see
that for spin-connections they have the gauge form,\footnote{\,Since the spin-connections have similar transformation behavior to that of the gauge fields, one should impose the Wess-Zumino consistency condition \cite{Zumino:1983rz} to obtain correct consistent anomalies. However, in $2d$ gauge theories, consistent anomalies have only two independent candidates \cite{Jo:1985} : $$\mathcal{A}_v=\int dx^2\ v_\alpha(c_1\partial^\mu A^\alpha_\mu+c_2\epsilon^{\mu\nu}\partial_\mu A^\alpha_\nu)\,.$$ In our sigma model, the left-handed chiral fermions only couple to $\omega_R$. Therefore, the anomalies are similar to those in the gauge theory, and will $not$ reduce to a purely topological term.}
\begin{eqnarray}
\delta_v\omega^a_{\ b\mu}=-\partial_\mu v^a_{\ b}-[\omega_\mu,  v]^a_{\ b}\,,
\label{spinconnT}
\end{eqnarray}
where the gauge function $v$, linear in parameters $\alpha^{i}$, depends on fields $\phi^{j}$.
Then the anomalies can be obtained by varying Eq.\,(\ref{Leff}),
\begin{eqnarray}
\mathcal{I}^{\,\rm total}_v\!\!\!&=&\!\!\delta_v\Gamma_{\rm eff}=\frac{1}{8\pi}\!\int \! d^2x\,\mbox{Tr}\ v\,\partial_L\omega_R\nonumber\\[1mm]
&=&\!\!\frac{1}{8\pi}\!\int \! d^2x\,\mbox{Tr}\,(v\,\partial^\mu\omega_\mu-v\,\epsilon^{\mu\nu}\partial_\mu\omega_\nu)\,.
\label{anomaly}
\end{eqnarray}
In Eq.\,(\ref{anomaly}), the first term can be removed by introducing a local counterterm
\begin{eqnarray}
S_{\rm c.t.}=-\frac{1}{16\pi}\int d^2x\mbox{Tr}\ \omega_\mu\omega^\mu\,.
\label{ct1}
\end{eqnarray}
This counterterm is in essence equivalent to adding heavy  Pauli-Villars (PV) fermions to ${\mathcal L}_f$,
\beq
\mathcal{L}_{\rm PV}=\frac{i}{2}H_{aL}\nabla_R H^a_L+\frac{i}{2}H_{aR}\nabla_L H^a_R+iMH_{aL}H^a_R\,,
\eeq
where $H^a_{L,R}$ are real Weyl-Majorana fermions, $$\nabla_{R,L}H^a_{L,R}\equiv\partial_{R,L}H^a_{L,R}+\omega^a_{\ bR,L}H^b_{L,R}\,,$$ and $M$ is the PV  mass. At the very end $M\to \infty$. One can check that, after integrating out the PV fermions $H^a_{L,R}$, one recovers  Eq. (\ref{ct1}).

The second term in Eq.\,(\ref{anomaly}) is purely topological. For simplicity, one can write it as a pulled-back form from ${\rm S}^{N-1}$ where we defined our sigma model by mapping $\phi:\Sigma\rightarrow {\rm S}^{N-1}$,
\beqn
\mathcal{I}_v&=&-\frac{1}{8\pi}\int_\Sigma d^2x\,\mbox{Tr}\,(v\epsilon^{\mu\nu}\partial_\mu\omega_\nu)=-\frac{1}{8\pi}\int_\Sigma \phi^*(\,\mbox{Tr}\,(vd\omega))\nonumber\\[2mm]
&=&-\frac{1}{8\pi}\int_{\phi(\Sigma)}\!\! \!\! \mbox{Tr}\,(vd\omega)\,.
\label{anomaly2}
\eeqn
The explicit expression for $\omega^a_{\ b}=\omega^a_{\ b\mu}dx^{\mu}$  can be calculated from Eqs.\,(\ref{8}) and (\ref{ome}),
\begin{equation}
\omega^a_{\ b}=\frac{2\phi^id\phi^j}{1+\phi^2}\,E^{\ a}_{ij\ b}\,, \qquad
E^{\ a}_{ij\ b}=\delta_i^{\ a}\delta_{jb}-\delta_j^{\ a}\delta_{ib}\,.
\label{Ospinconnection}
\end{equation}
Here the $E_{ij}$'s are the generators of the $\mathfrak{so}(N\!-\!1)$ Lie algebra in fundamental representation,  the holonomy group of ${\rm S}^{N-1}$ is ${\rm SO}(N\!-\!1)$.
Then variation of spin-connection $\omega$ with respect to $\alpha^{i}$ transformations of Eq.\,(\ref{isotranf})
has the form (\ref{spinconnT}) with $v$ given by
\begin{equation}
v^a_{\ b}=-\alpha^i\phi^jE^{\ a}_{ij\ b}\,.
\label{NLOnIsoT}
\end{equation}
Therefore the anomaly is given by  Eq.\,(\ref{anomaly2}) with $v$ from Eq.\,(\ref{NLOnIsoT}).

With $v^{a}_{\ b}$ being  $\phi$-dependent the integrand in (\ref{anomaly2}) does not look as a total derivative. However, it can be rewritten, using
integration by parts, as follows:
\begin{eqnarray}
\delta_\alpha\Gamma_{\rm eff}&=&\frac{1}{8\pi}\int_{\phi(S^2)}dv^{a}_{\ b}\wedge\omega^{b}_{\ a}=\frac{1}{8\pi}\int_{\phi(S^2)}\frac{2\alpha_i\phi_j}{1+\phi^2}\,d\phi^i\wedge d\phi^j\nonumber\\[3mm]
&=&\frac{1}{8\pi}\int_{\phi(S^2)}d\left[ \log(1+\phi^2)\,\alpha_id\phi^i\right].
\end{eqnarray}
Then we see that the variation is, in fact, an integral of a total derivative. Therefore, the local anomalies of isometries in the ${\rm O}(N)$ models vanish.

\section{\boldmath{${\rm CP}(N\!-\!1)$} Sigma Model}
\label{cpsm}

Our second example is the ${\rm CP}(N\!-\!1)={\rm SU}(N)/{\rm S}({\rm U}(N\!-\!1)\times {\rm U}(1))$ sigma model \cite{Witten: 1979}. The model involves $N$ complex fields $u^i$ ($i=1,2,...,N$) with the constraint $$\bar{u}_iu^i=1\,.$$ In addition we need to impose a local $\rm{U}(1)$
gauge invariance under
\beq
u^i\rightarrow{\rm e}^{i\alpha(x)}u^i\,.
\label{u1gauge}
\eeq
To this end one introduces an auxiliary vector field $A_\mu\,$, and the
  Lagrangian takes the form
\beq
\mathcal{L}_{b}=\frac{2}{g^2_0}(\partial_{\mu}+iA_\mu)\bar{u}_i(\partial^{\mu}-iA^\mu)u^i+\lambda(\bar{u}_i u^i-1)\,.
\label{LagCP}
\eeq

Similarly to the ${\rm O}(N)$ case, to pick up a patch we can chose a
``complex"axis, e.g.,  $u^N$. The isometries of the model fall into
two groups: linear transformations which do not transform $u^N$,
\begin{equation}
\delta_\epsilon u^i=\epsilon^{i\bar{j}}u_{\bar{j}}\,, \ \ \ \delta_{\epsilon}u^N=0\,; \quad i,\bar{j}=1,2,...,N-1\,,
\label{CPtranf}
\end{equation}
 and nonlinear ones which rotate $u^N$,
 \begin{equation}
\delta_{\beta}u^i=\beta^{i}u^N, \ \ \ \delta_{\beta}u^N=-\bar{\beta}_{i}u^i\,; \quad i=1,2,...,N-1\,.
\label{CPtranf1}
\end{equation}
In the above expressions, $\epsilon^{i\bar{j}}$ is an anti-Hermitian matrix and thus has $(N-1)^2$ real parameters while $\beta^i$ are $N-1$ complex parameters. The indices can be locally raised or lowered by $\delta^{i\bar{j}}$ or $\delta_{i\bar{j}}$. The total number of isometries is $N^2-1$ corresponding to ${\rm SU}(N)$ symmetries of the ${\rm CP}(N-1)$ model. Furthermore,  since $A_\mu$ is nondynamical, we can eliminate  it in favor of the $u^i$ fields,
\beq
A_\mu=-\frac{i}{2}(\bar{u}_i\partial_\mu u^i-\partial_\mu\bar{u}_iu^i)\,.
\label{gauge}
\eeq

Now, we can fix the gauge by condition ${\rm Im}\,u^N=0$, and solve the constraint by choosing a set of local coordinates $\{\phi^i, \ \bar{\phi}^{\bar{j}}\}$,
\beqn
\phi^i=\frac{u^i}{u^N}\,, \qquad  i=1,2,...,N-1\,.
\label{chart}
\eeqn
In terms of the new coordinates the isometry transformations of the model are
\beqn
\delta_\epsilon\phi^i&=&\epsilon^{i\bar{j}}\phi_{\bar{j}}\,;\nonumber\\[1mm]
\delta_\beta\phi^i&=&\beta^{i}\,,\nonumber\\[1mm]
\delta_{\bar{\beta}}\phi^i&=&(\bar{\beta}\phi)\phi^{i}\,.
\label{CPtranf2}
\eeqn

Parallelizing  our discussion of the ${\rm O}(N)$ model, we can write down the Lagrangian in terms of the fields $\phi^i, \ \bar{\phi}^{\bar{j}}$. We then supersymmetrize it to form a $\mathcal{N}=(0,2)$ $\rm{CP}(N\!-\!1)$ model by coupling complex \emph{left}-handed Weyl fermions $\psi^i\equiv \psi^i_L$,
\beq
\mathcal{L}=\mathcal{L}_b+\mathcal{L}_{f}=g_{i\bar{j}}\partial_{\mu}\bar{\phi}^{\bar{j}}\partial^{\mu}\phi^i
+g_{i\bar{j}}\,\bar{\psi}^{\bar{j}}i\gamma^{\mu}D_{\mu}\psi^i\,,
\label{L_NCP}
\eeq
where
\beqn
g_{i\bar{j}}=\frac{2}{g^2_0}\,\frac{(1+\bar{\phi}_i\phi^i)\delta_{i\bar{j}}-\bar{\phi}_i\phi_{\bar{j}}}{(1+\bar{\phi}_i\phi^i)^2}
\eeqn
is the standard Fubini-Study metric for ${\rm CP}(N\!-\!1)$.

To explore the isometry anomalies, one can introduce vielbeins as in the ${\rm O}(N)$ model, but the calculation is lengthy and tedious. We present the calculation details in Appendix. Here, instead, we will find a relation between the gauge anomaly in the gauged linear model and the isometry anomalies in the nonlinear formulation.

\vspace{2mm}


The full $\mathcal{N}=(0,2)$ ${\rm CP}(N-1)$ gauged model is obtained by adding $N$ complex left-handed fermions $\xi_L^{i}$ with constraints $u^i\bar{\xi}_{iL}=0$.  The corresponding Lagrangian takes the form
\beq
\mathcal{L}=\mathcal{L}_b+\frac{2}{g^2_0}\,\bar{\xi}_{Li}(i\partial_R+A_R)\xi^i_L+
\frac{2}{g^2_0}\left(\kappa_R\bar{\xi}_{iL}u^i+{\rm H.c.}\right),
\label{L_CP}
\eeq
where $\kappa_R$ is a Lagrange multiplier. The one-loop effective fermionic action following from the bi-angle diagram similar to Fig.\ \ref{Biangle} is
\beq
i\Gamma_{\rm eff}[A_R]=-\frac{iN}{8\pi}\int d^2x\ A_{R}\,\frac{\partial_L\partial_L}{\partial^2}\,A_{R}\,.
\label{CPeff}
\eeq
This action obviously suffers from a ${\rm U}(1)$ anomaly. Similarly to Eq.\,(\ref{anomaly}), this anomaly has longitudinal and topological parts. Since the anomaly in the longitudinal term is always cancelable by a counterterm, as Eq.\,(\ref{ct1}), we will focus on the topological part.

Keeping in mind that the   gauge transformation has the form
\beq
A_\mu \rightarrow A_\mu+\partial_\mu\alpha(x)
\eeq
we obtain the anomaly
\beq
\mathcal{A}_\alpha=-\frac{N}{4\pi}\int \alpha dA\,,
\eeq
where for simplicity we presented the anomaly via a one-form, $A=A_\mu dx^\mu$.

Now let us connect the  nonlinear isometry anomalies with the  gauge anomaly. To write the nonlinear sigma model, we need fix a gauge and choose a local chart to solve the constraints, see Eq.\,(\ref{chart}). Once the gauge and the chart are chosen, the isometries of rotation around $u^N$ are linear, while those rotating the $u^N$ axis have to be nonlinearly realized, see Eqs.\,(\ref{CPtranf}) and (\ref{CPtranf1}).

Equation (\ref{gauge}) implies that, since $A_\mu$ is isometry invariant, so is the fermion effective action
($\ref{CPeff}$). The  only  anomaly that exists in the gauged ${\rm CP}(N\!-\!1)$ formulation is the gauge anomaly. Then how can we have isometry anomalies produced? Notice that, within the fixed gauge, the  $u^N$ field must be real. However,  after rotations of $u^N$ this field becomes complex again. Therefore, to satisfy the reality condition for $u^N$,  the non-linear isometry transformations must be accompanied by a corresponding gauge transformation to offset the imaginary part of $u^N$.   This leads to the gauge anomaly, or equivalently, isometry anomalies in geometric formulation. We also verify that it is indeed the anomalies of the nonlinear isometries by straightforward calculation in Appendix.

Following the discussion above, we want to find a gauge parameter $\alpha$, such that $\delta_{\alpha}u^N+\delta_{\beta}u^N$ is real. Since
\beq
\delta_{\alpha}u^N+\delta_{\beta}u^N=i\alpha u^N-\bar{\beta}_i u^i
\eeq
where $i=1,2,...,N\!-\!1$, the reality condition is
\beq
i\alpha u^N-\bar{\beta}_i u^i=-i\alpha u^N-\beta^i \bar{u}_i\,.
\eeq
Therefore,  we can find $\alpha$ in terms of $\phi^i$ and $\bar{\phi}_i$, namely,
\beq
\alpha=\frac{i}{2}\,(\beta\bar{\phi}-\bar{\beta}\phi)\,.
\label{speGauge}
\eeq
Furthermore,  we rewrite the gauge field $A$ in terms of $\phi^i$ and $\bar{\phi}_i$ as well,
\beq
A=-\frac{i}{2}\,\frac{\bar{\phi}d\phi-d\bar{\phi}\phi}{1+\bar{\phi}\phi}\,,
\eeq
what gives for $dA$
\begin{equation}
dA=\frac{ig_{0}^{2}}{2}\, g_{i\bar j}\,d\phi^i\wedge d\bar{\phi}^{\bar{j}}\,.
\end{equation}
In this way obtain the nonlinear isometry anomalies,
\beqn
\mathcal{I}_\beta=\mathcal{A}_\alpha&\!\!=\!\!&\frac{i}{8\pi}\int(\bar{\beta}\phi-\beta\bar{\phi})\,\left[iN\,\frac{(1+\bar{\phi}\phi)\delta_{i\bar{j}}-\bar{\phi}_i\phi_{\bar{j}}}{(1+\bar{\phi}\phi)^2}\,d\phi^i\wedge d\bar{\phi}^{\bar{j}}\right]\nonumber\\[2mm]
&\!\!=\!\!&\frac{i}{8\pi}\int(\bar{\beta}\phi-\beta\bar{\phi})\,c_1\,,
\label{CPGLanomaly}
\eeqn
where $c_1$ is the first Chern class of ${\rm CP}(N\!-\!1)$.

In contradistinction with the ${\rm O}(N)$ sigma model (with the exception of ${\rm CP}(1)$, see below)  all other ${\rm CP}(N\!-\!1)$ sigma models suffer from the isometry anomalies which are neither a total derivative nor cancelable by adding local counter terms. For ${\rm CP}(1)$, the situation is identical to the ${\rm O}(3)$ model.

\subsection{\boldmath{${\rm CP}(1)$} is a special case}

We find from Eq.\,(\ref{CPGLanomaly}) it is indeed total derivative and consistent with previous discussion on ${\rm O}(N\!-\!1)$ model. The specialty that distinguish ${\rm CP}(1)$ from other ${\rm CP}(N\!-\!1)$ models  is its low dimension. Since ${\rm CP}(1)$ is geometrically a two dimensional sphere, locally we only have one $\phi$ and one $\bar{\phi}$ on one local chart. Equation (\ref{CPGLanomaly}) can be greatly simplified in this case and  written as an integral over total derivative:
\beqn
\mathcal{A}_{{\rm CP}(1)}=-\frac{N}{8\pi}\int\frac{\bar{\beta}\phi-\beta\bar{\phi}}{(1+\bar{\phi}\phi)^2}\,d\phi\wedge d\bar{\phi}=-\frac{N}{8\pi}\int d\left(\frac{\bar{\beta}d\phi+\beta d\bar{\phi}}{1+\bar{\phi}\phi}\right)\,.
\eeqn

On the other hand, globally ${\rm CP}(1)$ sigma model is known to have zero first Pontryagin class $p_1$, because it at most supports nonzero two-form while $p_1$ is an element in the fourth de Rham cohomology group. So far the local anomalies calculations are consistent with the global analysis of \cite{Moore:1984ws}.

In this section we found the relation between isometry anomalies $\mathcal{I}$ and gauge anomaly $\mathcal{A}$. The isometry anomalies in geometric formulation can be understood as gauge anomaly of a special gauge transformation, see Eq.\,(\ref{speGauge}). Following this clue, one can prospect the correspondence of holonomy anomaly versus arbitrary gauge anomaly, and further global anomaly versus ``large'' gauge anomaly, in geometric and gauge formulations respectively.

\subsection{A closer look at the correspondence between
 isometry and gauge anomalies}
\label{revisit cpn}

In this subsection, we want to discuss the correspondence between the isometry and gauge anomalies in a more rigorous mathematical way. It will also help us to apply these results in calculating the isometry anomalies in the general coset $G/H$ minimal sigma model in our subsequent work.

First, we want to rephrase the construction of ${\rm CP}(N\!-\!1)$ sigma model in the language of fiber bundle. The Lagrangian, Eq.\,(\ref{L_CP}) or (\ref{L_NCP}), is constructed through the famous Hopf fibration, see Fig.\,\ref{fiberation} below, by considering ${\rm CP}^{N-1}$ as the base space of the ${\rm U}(1)$ principal bundle of ${\rm S}^{\,2N-1}$, i.e.
\beq
\xymatrix{{\rm U}(1)\ar[r]^i & {\rm S}^{2N-1} \ar[r]^{\pi} &{\rm CP}(N\!-\!1)}\,.\nonumber
\eeq

\begin{figure}[t]
\begin{center}
\includegraphics[width=5.5cm]{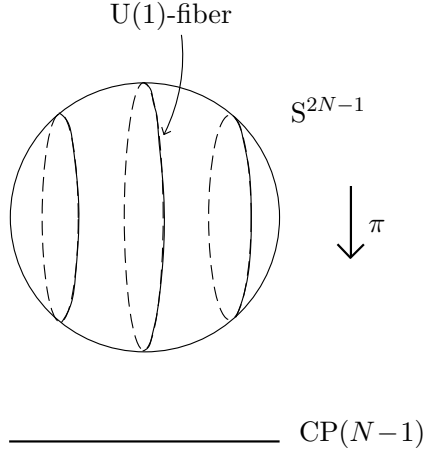}
\end{center}
\caption{\small The sphere denotes ${\rm S}^{2N-1}$, while the solid line below is for ${\rm CP}(N\!-\!1)$. The vertical circles are ${\rm U}(1)$-fibers, each of which is projected to a point on ${\rm CP}(N\!-\!1)$.}
\label{fiberation}
\end{figure}

\noindent
Equation (\ref{chart}) and the gauge condition that $u^N$ is fixed to be real actually assign a map, or, say, a local section, from a local chart $U_s\subset{\rm CP}^{N-1}$ to ${\rm S}^{2N-1}$,
\beqn
&&\xymatrix{s:\ U_s \ar[r] & {\rm S}^{2N-1}}\nonumber\\
&&\hspace{3mm} {\xymatrix{(\phi^i, \bar{\phi}_j) \ar@{|->}[r] & (u^i, u^N, \bar{u}_i, \bar{u}_N)}=\Big(\frac{\phi^i}{\rho}, \frac{1}{\rho}, \frac{\bar{\phi}_j}{\rho}, \frac{1}{\rho}\Big)}\ {\rm for}\ i, j=1, 2,..., N-1\,, \nonumber
\eeqn
with $$\rho=(1+\bar{\phi}_i\phi^i)^{1/2}\,.$$ Therefore it defines a local trivialization $\Phi$ of ${\rm S}^{2N-1}$, so that ${\rm S}^{\,2N-1}$ locally looks like a product space of $U_s\times {\rm U}(1)$
\beqn
&&\xymatrix{\Phi:\ U_s\times {\rm U}(1) \ar[r] & {\rm S}^{2N-1}}\nonumber\\
&&\hspace{8mm} {\xymatrix{(\phi^i, \bar{\phi}_j; e^{i\alpha}) \ar@{|->}[r] & s(\phi, \bar{\phi})e^{i\alpha}}\equiv\Big(\frac{\phi^i}{\rho}\,e^{i\alpha}, \frac{1}{\rho}\,e^{i\alpha}, \frac{\bar{\phi}_j}{\rho}\,e^{-i\alpha}, \frac{1}{\rho}\,e^{-i\alpha}\Big)}.
\label{trivialization}
\eeqn

It is easy to see that the ${\rm U}(1)$-action on fiber $\pi^{-1}(\phi, \bar{\phi})$, is just the gauge transformation (\ref{u1gauge}). We want to point out that, the ${\rm U}(1)$-gauge $A_\mu$ in Eq.\,(\ref{gauge}) is exactly a choice of \emph{connection} $1$-form defined on the bundle ${\rm S}^{\,2N-1}$.

To see this, one needs to recall how to define a ${\rm U}(1)$-connection on the principal bundle ${\rm S}^{\,2N-1}$. First, the ${\rm U}(1)$-action moves any point $p\in {\rm S}^{2N-1}$ along the fiber, which defines a one-dimensional subspace of the tangent space $\T_p{\rm S}^{2N-1}$, called vertical space $V_p$ (see the tangential direction of the vertical circles in Fig.\ \ref{fiberation}). The corresponding tangent vector $\sigma_p$ spanning $V_p$ is called the fundamental vector, and is given by the trivialization (\ref{trivialization}) as
$$
\sigma_p=iu^i\frac{\partial}{\partial u^i}-i\bar{u}_i\frac{\partial}{\partial\bar{u}_i},\ {\rm for}\ i=1,2,...,N
$$
subject to the constraint $\bar{u}_iu^i=1$. Now we are about to assign a ${\rm U}(1)$ connection on ${\rm S}^{2N-1}$, or  a $2N-2$ dimensional horizontal subspace $H$, so that the tangent space of bundle ${\rm S}^{2N-1}$ can be decomposed as direct sum of horizontal and vertical spaces:
$$\T_pS^{\,2N-1}=H_p\oplus V_p, \ {\rm for}\ p\in {\rm S}^{2N-1}.$$

Equivalently, using a more familiar language, $H$ is determined by a $1$-form
$$\tilde A\in \Omega^1({\rm S}^{2N-1})$$ globally defined on ${\rm S}^{2N-1}$, so that
$$H_p={\rm Span}\{X_p\in \T_p{\rm S}^{2N-1}\vert \tilde A_p(X_p)=0\}={\rm ker}\tilde A_p$$
with $\tilde A$ satisfying
\beq
\tilde A_p(\sigma_p)=1\ \ {\rm and} \ \ R^{\ast}_\alpha\tilde A_p=\tilde A_{pe^{i\alpha}}\ {\rm for}\ p\in {\rm S}^{2N-1}\,.
\label{Acond.}
\eeq
In the second equation $R^\ast_\alpha$ is the pullback induced by the ${\rm U}(1)$ action on fibers, which guarantees the equivariance of $\tilde A$.

Generically there are various ways to choose the horizontal space $H_p$ corresponding to different connection $1$-forms $\tilde A$. However, for ${\rm CP}^{N-1}$ as the quotient space of ${\rm S}^{2N-1}$ by ${\rm U}(1)$, the projection map is a Riemann submersion once we assign the standard round metric $\tilde g$ on $S^{\,2N-1}$. Its tangent space at $\pi(p)$, i.e. $\T_{\pi(p)}{\rm CP}^{N-1}=\pi_\ast H_p$, is an orthogonal complement to $V_p$.

The metric $\tilde g$ defined on the standard sphere ${\rm S}^{2N-1}$, with the coordinates $(u^i, \bar{u}_i)$, is  given by
$$
\tilde g=\frac{1}{2}(d\bar{u}_i\otimes du^i+du^i\otimes d\bar{u}_i),\ {\rm with}\ \bar{u_i}u^i=1\,.
$$
Since we choose the horizontal space $H_p=V^{\perp}_p=({\rm Span}\{\sigma_p\})^\perp$, the connection $1$-form $\tilde A$ is thereby proportional to $\tilde g(\sigma_p)$,
$$
\tilde A\sim\tilde g(\sigma_p)=-i\left(\bar{u}_idu^i-d\bar{u}_iu^i\right).
$$
To meet the condition (\ref{Acond.}), we fix the coefficient of $\tilde A$ as
$$
\tilde A=-\frac{i}{2}\left(\bar{u}_idu^i-d\bar{u}_iu^i\right).
$$
One can see that this is just Eq.\,(\ref{gauge}).

Finally, the connection $\tilde A$ is pulled back from ${\rm S}^{2N-1}$ to the local chart $U_s$ of ${\rm CP}^{N-1}$ by section $s$. i.e.
\beqn
&&\xymatrix{s^{\ast}: \Omega^1({\rm S}^{2N-1})\ar[r] & \Omega^1(U_s)}\,,
\nonumber\\[1mm]
&&\hspace{6mm} {\xymatrix{\tilde A \ar@{|->}[r] & A}=s^{\ast}\tilde A=-\frac{i}{2}\frac{\bar{\phi}d\phi-d\bar{\phi}\phi}{1+\bar{\phi}\phi}}\,.
\eeqn

Once we assign a new section $$s^\prime: U_{s^\prime}\rightarrow {\rm S}^{2N-1}\,,$$ it is clear that, on the intersection $U_s\cap U_{s^\prime}$, any two points on one and the same fiber mapped by $s$ and $s^\prime$ are related by a ${\rm U}(1)$ action, i.e.
$$s^\prime(\bar{\phi}, \phi)=s(\bar{\phi}, \phi)e^{i\alpha(\bar{\phi}, \phi)},\ {\rm for}\ (\bar{\phi}, \phi)\in U_s\cap U_{s^\prime}.$$
Similarly, $s^\prime$ will also pullback the connection $\tilde A$ to $A^\prime=s^{\prime\ast}\tilde A$. Moreover, $A^\prime$ and $A$ are related by our familiar ${\rm U}(1)$-gauge transformation
$$
A^\prime=A+d\alpha\,.
$$

Based on the discussion above, the Lagrangian, Eq.\,(\ref{L_CP}) and Eq.\,(\ref{L_NCP}), could be interpreted as ${\rm CP}^{N-1}$ model constructed on the bundle ${\rm S}^{2N-1}$ or a local patch $U_s\subset {\rm CP}^{N-1}$. Now, when we consider an isometry transformation $f$ on a local patch, say, $U_s$, the isometry will induce a change of $U_s$ and thus a gauge transformation of $A$. Therefore when we calculate the isometry anomalies of ${\rm CP}^{N-1}$ localized on $U_s$, they are naturally associated to the gauge anomalies of $A$ locally defined on $U_s$.

To address the idea in a rigorous manner, we need the concept of the bundle isomorphism. A bundle isomorphism is a $1-1$ bundle map $F$ such that the following diagram commutes:
\beq
\xymatrix{&{\rm S}^{2N-1} \ \ar_{\pi}[d] \ar^{F}[r] &{\rm S}^{2N-1} \ \ar^{\pi}[d]\nonumber
\\&{\rm CP}^{N-1} \ \ar^{f}[r] &{\rm CP}^{N-1}}\nonumber
\eeq
where $f$ is an induced isomorphism on base space ${\rm CP}^{N-1}$, and $F$ satisfies the equivariant condition:
\beq
 F(pe^{i\alpha})=F(p)e^{i\alpha}\,.
 \label{bundle equiv}
\eeq

As for the isometry transformations, one needs to consider the corresponding isometric bundle isomorphism, i.e. isomorphisms preserving given metric $\tilde g$ on the bundle ${\rm S}^{2N-1}$ and satisfying equivariance Eq.\,(\ref{bundle equiv}). For the bundle ${\rm S}^{2N-1}$, there are $2N^2-2N$ isometries as we discussed in Sec.\ \ref{On}, while only the transformations (\ref{CPtranf}) satisfy the condition (\ref{bundle equiv}) and induce the isometries on ${\rm CP}^{N-1}$.

Now we are interested in the transformation of the connection $A$ with respect to isometric bundle morphisms. Generically a bundle isomorphism $F$ will ``pushforward" the horizontal space $H$ to
$$
H^F\equiv F_{\ast}H\,.
$$
Therefore the corresponding connection $1$-form for $H^F$ is
$$
\tilde A^F=(F^{-1})^{\ast}\tilde A\,.
$$
We want to calculate the difference of $A^F$ from $A$ pulled back to the base space ${\rm CP}^{N-1}$, e.g. at the point $b\in U_s\subset {\rm CP}^{N-1}$. Note that the isometric bundle morphism $F$, i.e. Eq.\,(\ref{CPtranf}), also induces an isometric morphism $f$ on ${\rm CP}^{N-1}$, see Eq.\,(\ref{CPtranf2}). Isomorphism $f$ moves the point $b=(\phi, \bar{\phi})$ to $c=f(b)$, located on a different fiber $\pi^{-1}(c)$. One thus needs to further pullback the connection by $f^\ast$ to compare their difference, see the commuting diagram below,
\beq
\xymatrix{& {\rm S}^{2N-1} & {\rm S}^{2N-1}\ar_{F^{-1}}[l]\nonumber
\\\ \ & U_{s^\prime} \ar_{s^\prime}[u] \ar^{f}[r] & U_s \ar_s[u]\nonumber}
\xymatrix{&\Omega^1({\rm S}^{2N-1}) \ \ar_{s^{\prime\ast}}[d] \ar^{(F^{-1})^\ast}[r] &\Omega^1({\rm S}^{2N-1}) \ \ar^{s\ast}[d]\nonumber
\\\ \ &\Omega^1(U_{s^\prime}) \  &\Omega^1(U_s)\ar_{f^\ast}[l]}\nonumber
\label{isodiagram}
\eeq
The variance of connection
$$f^{\ast}\circ s^\ast \tilde A^F-s^\ast \tilde A=(F^{-1}\circ s\circ f)^\ast \tilde A-s^\ast \tilde A$$
respect to point $b=(\phi, \bar{\phi})$, will be considered. However, the combination of maps $s^\prime \equiv F^{-1}\circ s\circ f$ defines another section,\footnote{\,It is the the pulled-back section of $s$ by $F$, and hence depends on the bundle map.}
$$s^\prime: U_{s^\prime}\rightarrow {\rm S}^{2N-1}\,.$$ Therefore one has
\beq
s^\prime(b)=s(b)e^{i\alpha(b)},\ \ \forall\ b\in U_s\cap U_{s\prime}\,,
\label{ss'}
\eeq
and therefore
\beq
s^{\prime\ast}\tilde A-s^\ast\tilde A=A^\prime-A=d\alpha\,.
\label{gaugetransf}
\eeq
Given Eq.\,(\ref{CPtranf}) and Eq.\,(\ref{CPtranf2}) for infinitesimal versions of $F$ and $f$, we can calculate the infinitesimal transformation of Eq.(\ref{gaugetransf}). We only consider transformations corresponding to the parameters $\beta$ and $\bar{\beta}$. Since
$$
e^{i\alpha(\phi,\bar{\phi})}\sim 1+i\alpha(\phi,\bar{\phi}, \beta, \bar{\beta})\,,
$$
 the infinitesimal transformation of Eq.\,(\ref{gaugetransf}) and Eqs.\,(\ref{CPtranf}) and (\ref{CPtranf2}) lead us back to Eq.\,(\ref{ss'}). After a short calculation, we obtain
$$
\alpha(\phi,\bar{\phi}, \beta, \bar{\beta})=\frac{i}{2}(\beta\bar{\phi}-\bar{\beta}\phi)\,,
$$
which coincides with the previous result (\ref{speGauge}).

So far we revisited the anomaly of the ${\rm CP}(N\!-\!1)$ sigma model. The lesson one can draw is that the nonlinear formalism Lagrangian, see Eq.\,(\ref{L_NCP}), is defined on the local patch $U_s$ of ${\rm CP}(N\!-\!1)$. An isometric transformation $f$, or $F$ on the bundle will result in a change of the local patch to $U_{s^\prime}$, or equivalently a change of the local section to $s^\prime$. Therefore the pulled-back connection, or the gauge field (\ref{gauge}), will transform as in Eq.(\ref{gaugetransf}). If there are chiral fermions coupling to the gauge field nontrivially, there must be anomalies produced. In this sense, these anomalies  measure the failure of bundle reparametrization from the section $s$ to $s^\prime$ induced by isometric transformation.

\section{Dual formalism for the \boldmath{$ {\rm O}(N)$} Model}
\label{dual}

In Sec.\ \ref{cpsm} we demonstrated that the isometry anomalies in the nonlinear realization of the
${\rm CP}(N-1)$ is in one-to-one correspondence with the ${\rm U}(1)$ anomaly in its gauged linear formulation.
This section is motivated by further consistency checks of the gauge versus isometry anomalies. Another motivation
is the large-$N$ argument regarding the gauge anomaly in the linear gauged sigma models.

A crucial difference between the ${\rm O}(N)$ and ${\rm CP}(N\!-\!1)$ sigma models is that the latter has a ${\rm U}(1)$ gauge field, and eventually suffers from the ${\rm U}(1)$ anomaly. At the same time, the ${\rm O}(N )$ sigma model  has no gauge redundancy and therefore is   expected to  have no isometry anomalies after the passage to its nonlinear formulation.

In Sec.\ \ref{On} we considered the ${\rm O}(N)$ model (which can also be called the $S^{N-1}$ model) using the realization of the target space in terms of $N$ real fields $n^i$ with the constraint (\ref{c1}). The real Grassmann model prompts us a dual form  of the ${\rm O}(N)$ model.

The same target space,  ${\rm S}^{N-1}$, can be implemented as follows.
Consider real bosonic matrix fields
\beq
N^\alpha_{\ a}\,,\qquad \alpha=1,2,...,N\,,\qquad a=1,2,...,N-1\,,
\label{41}
\eeq
and gauge the ${\rm SO}(N-1)$ symmetry. The index $\alpha$ in (\ref{41}) will play the role
of the ``color" index of the gauged group ${\rm SO}(N-1)$. The index $a$ is the ``flavor"  index of global ${\rm SO}(N)$ symmetry.
Then we add a  constraint
\beq
(N^T)^{a}_{\ \ \alpha} N^{\alpha}_{\ b}=\delta^{a}_{\ b}\,.
\eeq
We also add left-handed fermions $\psi^\alpha_{La}$ with appropriate constraints to
 supersymmetrize the model. In this way we arrive at the Lagrangian
\beqn
&&\mathcal{L}=\frac{1}{2g_0^2}{\rm Tr}[(D^\mu N)^TD_\mu N+i\bar{\psi_L}D_R\psi_L]\nonumber\,,\\[3mm]
&&N^{Ta}_{\ \ \alpha} N^{\alpha}_{\ b}=\delta^{a}_{\ b}\,,\qquad (N^{T})^{a}_{\ \ \alpha}\psi^\alpha_{Lb}=0\,.
\label{43}
\eeqn
where
\beq
(D_\mu N)^{\alpha}_{\ a}=\partial_\mu N^{\alpha}_{\ a}-N^{\alpha}_{\ b}A^b_{\mu a}\,,
\eeq
 and the matrix fields $A^b_{\mu a}$ are the ${\rm SO}(N\!-\!1)$ gauge fields. As previously mentioned, the above gauge fields are nondynamical and can be eliminated in favor  of the $N$ fields,
\beq
A^a_{\mu b}=\frac{1}{2}\left(N^T\partial_\mu N-\partial_\mu N^T\!\cdot N\right)^a_{\ b}\,.
\eeq

Similarly to Eqs. \,(\ref{gauge}) and (\ref{chart}) in the ${\rm CP}(N\!-\!1)$ model, one can fix an ${\rm SO}(N\!-\!1)$ gauge, and choose local charts to write down a nonlinear sigma model. For example, we treat
\beq
N^\alpha_{\ a}=
\left(
  \begin{array}{c}
    V^i_{\ a} \\
    \rho_a\\
  \end{array}
\right),\qquad a,i=1,2,...,N-1\,,
\label{splitting}
\eeq
where  $\rho_a\equiv N^N_{\ a}\, $ is an additional row vector.

Now,
we fix the gauge in such a way  that $V^i_{\ a}$ becomes symmetric real matrix. Then we define a local chart,
\beq
\phi_i=\rho_a\left(\frac{1}{1+V}\right)^{a}_{\ i}\,.
\eeq
After solving the constraint, one obtains
\beqn
&&V^i_{\ a}=\left(\delta^i_{\ j}-\frac{2}{1+\phi^2}\phi^i\phi_j\right)\delta^j_{\ a}\,,\nonumber\\[3mm]
&&\rho_a=\frac{2\phi_i}{1+\phi^2}\delta^i_{\ a}\,,\nonumber\\[3mm]
&&A^a_{\mu b}=\frac{2\phi^i\partial_\mu\phi^j}{1+\phi^2}E^{\ a}_{ij\ b}\,.
\label{DualOngauge}
\eeqn
The generators $E^{\ a}_{ij\ b}$ were defined in Eq.\,(\ref{Ospinconnection}).

One can easily convince oneself that the gauge fields $A_\mu$ are just spin connections $\omega_\mu$ in nonlinear
formulation of the ${\rm S}^{N-1}$ model, see Eq.\,(\ref{Ospinconnection}). Thus, the nonlinear Lagrangian following from
(\ref{43}) after gauge fixing  is in fact identical to that presented in Eq.\,(\ref{OnLagrangian}).

At the perturbative level, the gauge anomalies in the present section and the isometry anomalies
in Sec. \ref{On}
will  match each other too. We will discuss only  those isometry transformations that involve an interplay between $V^i_{\ a}$ and $\rho_a$,
\beqn
\delta_\alpha V^i_{\ a}=\alpha^i\rho_a\,, \qquad   \delta_\alpha\rho_a=-\alpha_iV^i_{\ a}\,,
\label{49}
\eeqn
since they would induce gauge anomalies for the fixed gauge, see the remark after Eq.~(\ref{splitting}). To keep the matrix $V^i_{\ a}$ symmetric, a gauge transformation must   accompany (\ref{49}), namely,
\beq
\delta_\lambda V^i_{\ a}=V^i_{\ b}\lambda^b_{\ a},\ \ {\rm with}\ \lambda^T=-\lambda\,.
\eeq
Solving equation
\beq
\delta_{\alpha+\lambda}V=(\delta_{\alpha+\lambda}V)^T,
\eeq
we arrive at
\beq
\lambda^a_{\ b}=v^a_{\ b}\,,
\label{OngaugeT}
\eeq
where the matrix $v^a_{\ b}$ is given in  Eq.~(\ref{NLOnIsoT}). Therefore, the induced gauge anomalies are
\beq
\mathcal{A}_\lambda=-\frac{1}{8\pi}\int {\rm Tr}\,(\lambda dA)\,.
\eeq
Equations (\ref{DualOngauge}) and (\ref{OngaugeT}) show that $\mathcal{A}_\lambda$ is just the nonlinear isometry anomalies, the same as  in Eq.\,(\ref{anomaly2}). The theory can be ``mended" just in the same way as it was
discussed in Sec.\ \ref{On}.

\section{Conclusions}
\label{concl}

Two-dimensional chiral sigma models with various degrees of supersymmetry present an excellent theoretical laboratory.
While the $(2,2) $ models were thoroughly explored in the 1980s,  chiral models received much less attention.
Recently non-minimal chiral models reappeared in the focus of theorists' attention because of their special role as
world-sheet models on topological vortex solutions supported in certain four-dimensional ${\cal N}=1$ Yang-Mills theories. This fact naturally raised interest to the minimal chiral models.

In this paper the minimal chiral two-dimensional models are revisited. We demonstrate that the Moore-Nelson consistency condition \cite{Moore:1984ws} revealing a global anomaly in ${\rm CP}(N\!-\!1)$  (with $N\!>\!2$) due to a nontrivial first Pontryagin class  is in one-to-one correspondence with the local anomalies in isometries. These latter anomalies are generated by fermion loop diagrams which we explicitly calculate.

At the same time the first Pontryagin class in the
${\rm O}(N)$ models vanishes \cite{Moore:1984ws} and, thus, these models are globally self-consistent. We show that the divergence of the isometry currents in these models is anomaly free. Thus, there are no obstructions to quantizing the minimal ${\mathcal N}=(0,1)$ models with the $S^{N-1}= {\rm SO}(N)/{\rm SO}(N\!-\!1)$ target space. ${\rm CP}(1)$ is self-consistent and presents an exceptional case from the ${\rm CP}(N\!-\!1)$ series: both
the first Pontryagin class vanishes and the local anomalies are absent too. We discuss a relation between the geometric and gauged formulations of the ${\rm CP}(N\!-\!1)$ models. From the standpoint of the principal fiber bundle, the isometry anomalies on a local patch just reflect the failure of gauge invariance of the theories in passing from one local patch to another. Therefore it  relates the local anomalies to global topological criteria
\cite{Manohar:1984zj,Moore:1984ws}.

 In our subsequent work, we will follow this clue to discuss anomalies in general minimal $G/H$ sigma models in both local and global aspects. The obvious distinction between the ${\rm O}(N)$ and ${\rm CP}(N\!-\!1)$ target spaces is the fact that in the first case the factor $H$ is a simple group, while in the second case it is a product two factors, ${\rm SU}(N\!-\!1)\times {\rm U}(1)$. One can conjecture that the non-simple character of $H$ is behind
emergence of anomalies, for the simple $H$ group the first Pontryagin class of $G/H$ vanishes. We will address this issue in \cite{CCSV3}.

\vspace{-4mm}

\section*{Acknowledgments}

We are grateful to N. Nekrasov for useful discussions. X.C. thanks the Max-Planck Institute for Mathematics in Bonn for hospitality. The  work of M.S. is supported in part by DOE grant DE-SC0011842. X.C. is supported by the Dorothea-Schl\"{o}zer Fellowship at the Georg-August Universit\"{a}t G\"{o}ttingen.

\newpage
%
%
%
\section*{Appendix: Vielbeins and Anomalies in \boldmath{${\rm CP}(N\!-\!1)$}}

\renewcommand{\theequation}{A.\arabic{equation}}
\setcounter{equation}{0}

The Fubini-Study metric $g_{i\bar{j}}$ on ${\rm CP}(N\!-\!1)$ is
\begin{eqnarray}
g_{i\bar{j}}=\frac{(1+\bar{\phi}_i\phi^i)\delta_{i\bar{j}}-\bar{\phi}_i\phi_{\bar{j}}}{(1+\bar{\phi}_i\phi^i)^2}\,.
\end{eqnarray}
The indices of charts $\{\phi^i, \bar{\phi}^{\bar{j}}\}$ locally are raised or lowered by $\delta^{i\bar{j}}$ or $\delta_{i\bar{j}}$. To explicitly find vielbein of the metric, it is convenient to define
\begin{eqnarray}
&&r^2\equiv \bar{\phi}_i\phi^i\,,
\quad\rho^2\equiv 1+r^2\,,
\nonumber\\[2mm]
&&P_{i\bar{j}}\equiv\delta_{i\bar{j}}-\frac{\bar{\phi}_i\phi_{\bar{j}}}{r^2}\,,
\nonumber\\[2mm]
&&Q_{i\bar{j}}\equiv\frac{\bar{\phi}_i\phi_{\bar{j}}}{r^2}\,,
\end{eqnarray}
one can easily check the following properties:
\begin{eqnarray}
&&\delta_{i\bar{j}}=P_{i\bar{j}}+Q_{i\bar{j}}\,,
\nonumber\\[2mm]
&&P_{i\bar{j}}\bar{\phi}^{\bar{j}}=P_{i\bar{j}}\phi^i=0\,,
\nonumber\\[2mm]
&&Q_{i\bar{j}}\bar{\phi}^{\bar{j}}=\bar{\phi}_i, \ \ Q_{i\bar{j}}\phi^i=\phi_{\bar{j}}\,,
\nonumber\\[2mm]
&&P^2=P, \ \ Q^2=Q, \ \ PQ=QP=0\,.
\end{eqnarray}
As a result, the metric and vielbein could be written as
\begin{eqnarray}
&&g_{i\bar{j}}=\frac{1}{\rho^2}(P_{i\bar{j}}+\frac{1}{\rho^2}Q_{i\bar{j}}), \ \ g^{i\bar{j}}=\rho^2(P^{i\bar{j}}+\rho^2Q^{i\bar{j}})\,,
\nonumber\\[2mm]
&&e^{a}_{\ i}=\frac{1}{\rho}(P^a_{\ i}+\frac{1}{\rho}Q^a_{\ i}), \ \ e^{i}_{\ a}=\rho(P^i_{\ a}+\rho Q^i_{\ a})\,,
\nonumber\\[2mm]
&&e^{\ \bar{b}}_{\bar{j}}=\frac{1}{\rho}(P^{\ \bar{b}}_{\bar{j}}+\frac{1}{\rho}Q^{\ \bar{b}}_{\bar{j}}), \ \ e_{\bar{b}}^{\ \bar{j}}=\rho(P_{\bar{b}}^{\ \bar{j}}+\rho Q_{\bar{b}}^{\ \bar{j}})\,,
\nonumber\\[2mm]
&&e^{a}_{\ i}e^{i}_{\ b}=\delta^{a}_{\ b}, \ \ e_{\bar{a}}^{\ \bar{j}}e^{\ \bar{b}}_{\bar{j}}=\delta_{\bar{a}}^{\ \bar{b}}, \ \ \delta_{a\bar{b}}e^{a}_{\ i}e^{\ \bar{b}}_{\bar{j}}=g_{i\bar{j}}\,.
\label{CPviel}
\end{eqnarray}
Similarly to the ${\rm O}(N\!-\!1)$ model, the symbols $\delta_{a\bar{b}}$ or $\delta^{a\bar{b}}$ are used to lower or raise frame indices $\{a, \bar{b}, ....\}$. Since ${\rm CP}(N\!-\!1)$ are the
K\"{a}hler manifolds, there are two sets of vielbein, and correspondingly two sets of spin-connections one-form on the frame bundles ${\rm Hol}^{(1,0)}$ and ${\rm Hol}^{(0,1)}$,
\begin{eqnarray}
&&\omega^a_{\ b}=\omega^a_{\ bi}d\phi^i=e^a_{\ j}D_ie^j_{\ b}\ d\phi^i\,,
\nonumber\\[2mm]
&&\bar{\omega}^{\ \bar{a}}_{\bar{b}}=\bar{\omega}_{\bar{b}\ \ \bar{j}}^{\ \bar{a}}d\bar{\phi}^{\bar{j}}=D_{\bar{j}}e_{\bar{b}}^{\ \bar{i}}\ e_{\bar{i}}^{\ \bar{a}}d\bar{\phi}^{\bar{j}}\,,
\nonumber\\[2mm]
&&\omega^{\dagger}=\bar{\omega}\,.
\label{CPw}
\end{eqnarray}

Redefining $\psi^{a}=e^a_{\ i}\psi^i$, one can present the fermionic part of the
${\rm CP}(N-1)$ Lagrangian  as
\begin{eqnarray}
&ig_{i\bar{j}}\bar{\psi}^{\bar{j}}\gamma^{\mu}(\partial_{\mu}\psi^i+\Gamma^{i}_{jk}\partial_{\mu}\phi^j\psi^{k}
)=i\bar{\psi}^{\bar{a}}\gamma^{\mu}(\partial_{\mu}\delta_{\bar{a}b}+\Omega_{\bar{a}b\mu})\psi^b\,,&\\[2mm]
&\Omega_{\bar{a}b\mu}=\omega_{\bar{a}bi}\partial_{\mu}\phi^i-\bar{\omega}_{\bar{a}b\bar{j}}\partial_{\mu}\bar{\phi}^{\bar{j}}\,,&
\end{eqnarray}
where $\Omega$ is the pulled-back connections from frame bundle ${\rm Hol}^{(1,0)}\oplus {\rm Hol}^{(0,1)}$ of \mbox{${\rm CP}(N\!-\!1)$}. Identically to the discussion of ${\rm O}(N\!-\!1)$ models, one can evaluate linear and non-linear isometry transformations on connection $\Omega$, and imposes the Wess-Zumino consistency condition to find consistent anomalies. However the calculation are much more cumbersome than ${\rm O}(N\!-\!1)$ case. The details will be neglected, only main results are listed.

Firstly, the number of isometries of ${\rm CP}(N\!-\!1)$ are $N^2\!-\!1=(N\!-\!1)^2+2(N\!-\!1)$, in which there are $(N\!-\!1)^2$ linear symmetries corresponding to ${\rm U}(N\!-\!1)$-rotations of fields $\{\phi^i, \bar{\phi}^{\bar{j}}\}$. It also implies the holonomy group of ${\rm CP}(N\!-\!1)$ is ${\rm U}(N\!-\!1)$. The rest of $2N\!-\!2$ symmetries are non-linearly realized,
\begin{eqnarray}
\delta_\epsilon&=&\epsilon^{i\bar{j}}(\phi_{\bar{j}}\frac{\delta}{\delta\phi^i}-\bar{\phi}_{i}\frac{\delta}{\delta\bar{\phi}^{\bar{j}}})\,,
\nonumber\\[2mm]
\delta_\beta&=&\beta^{i}\frac{\delta}{\delta\phi^i}+(\beta\bar{\phi})\bar{\phi}^{\bar{j}}\frac{\delta}{\delta\bar{\phi}^{\bar{j}}}\,,
\nonumber\\[2mm]
\delta_{\bar{\beta}}&=&\bar{\beta}^{\bar{j}}\frac{\delta}{\delta\bar{\phi}^{\bar{j}}}+(\bar{\beta}\phi)\phi^{i}\frac{\delta}{\delta\phi^i}\,.
\end{eqnarray}
Further, we calculate the variation of spin-connection $\Omega$. According to the experience from ${\rm O}(N\!-\!1)$, it is not curious that linear symmetries give no anomalies to effective Lagrangian. Therefore only non-linear symmetries are considered as below. Since $\Omega$ is anti-Hermitian, one can only evaluate $\delta_\beta\Omega$, and take hermitian conjugation to get $\delta_{\bar{\beta}}\Omega$. After explicit calculations following Eq.(\ref{CPviel}) and (\ref{CPw}) we arrive at
\begin{eqnarray}
\omega^a_{\ b}&=&\Big(-\frac{1}{2\rho^2}\bar{\phi}_i\delta^a_{\ b}-\frac{1}{2\rho^2}\bar{\phi}_iQ^a_{\ b}-\frac{\rho-1}{\rho r^2}\bar{\phi}_bP^a_{\ i}\Big)d\phi^i
\nonumber\\[2mm]
&=&\Big(-\frac{1}{2\rho^2}\bar{\phi}_i\delta^a_{\ b}-\frac{\rho-1}{\rho r^2}\bar{\phi}_b\delta^a_{\ i}+\frac{1}{2}\,\frac{(\rho-1)^{2}}{\rho^{2} r^4}\,\bar{\phi}_i\phi^a\bar{\phi}_b\Big)d\phi^i
\nonumber\\[2mm]
&\equiv&\big[-G(r^2)\bar{\phi}_i\delta^a_{\ b}-F(r^2)\bar{\phi}_b\delta^a_{\ i}+\frac{1}{2}\,F^2(r^2)\bar{\phi}_i\phi^a\bar{\phi}_b\big]d\phi^i\,,
\nonumber\\[2mm]
\bar{\omega}^{\ \bar{a}}_{\bar{b}}&=&\big[-G(r^2)\phi_{\bar{j}}\delta^{\ \bar{a}}_{\bar{b}}-F(r^2)\phi_{\bar{b}}\delta^{\ \bar{a}}_{\bar{j}}+\frac{1}{2}F^2(r^2)\phi_{\bar{j}}\phi_{\bar{b}}\bar{\phi}^{\bar{a}}\big]d\bar{\phi}^{\bar{j}}\,,
\end{eqnarray}
where real functions $G$ and $F$ are defined as
\begin{eqnarray}
G(r^2)\equiv\frac{1}{2\rho^2}\,, \ \ \ \ \ \ F(r^2)\equiv\frac{\rho-1}{\rho r^2}\,.
\end{eqnarray}
Varying $\Omega^a_{\ b}=\omega^a_{\ b}-\bar{\omega}^a_{\ b}$, one must have
\begin{eqnarray}
\delta_{\beta}\Omega^a_{\ b}=-dv^{\ a}_{\beta\ b}-[\Omega, \ v_\beta]^a_{\ b}\,.
\label{wtransf}
\end{eqnarray}
To find $v^{\ a}_{\beta\ b}$ in  the easiest way one can consider variation of the torsion equation on ${\rm CP}(N\!-\!1)$. Since there is no torsion on ${\rm CP}(N\!-\!1)$, one has
\begin{eqnarray}
de^a+\Omega^a_{\ b}\wedge e^b=0,
\end{eqnarray}
where $e^a=e^a_{\ i}d\phi^i$ is frame one-form. Acting $\delta_{\beta}$ on both sides, one can obtain
 Eq.(\ref{wtransf}) if
\begin{eqnarray}
\delta_{\beta}e^a=v^{\ a}_{\beta\ b}e^b\,.
\end{eqnarray}
Explicitly calculating $\delta_{\beta}e^a$, we derive $v^{\ a}_{\beta\ b}$,
\begin{eqnarray}
v^{\ a}_{\beta\ b}&=&-\frac{\beta\bar{\phi}}{2}\delta^a_{\ b}-\frac{\beta\bar{\phi}}{2}Q^a_{\ b}-\frac{\rho-1}{r^2}\bar{\phi}_bP^a_{\ i}\beta^i\nonumber\\[2mm]
&=&-\frac{\beta\bar{\phi}}{2}\delta^a_{\ b}-\rho F\beta^a\bar{\phi}_b-\frac{\beta\bar{\phi}}{2}\rho^2F^2\phi^a\bar{\phi}_b\,.
\end{eqnarray}

Now the non-linear isometry anomalies of ${\rm CP}(N\!-\!1)$ can be assembled together by using the
Wess-Zumino consistency condition. Similar to the ${\rm O}(N\!-\!1)$ case, anomalies with respect to parameter $\beta$ are
\begin{eqnarray}
\delta_{\beta}\Gamma_{eff}&=&-\frac{1}{4\pi}\int_{\phi(S^2)}v^{\ a}_{\beta\ b}d\Omega^{b}_{\ a}
\label{CPNLanomaly}\\
&=&-\frac{1}{4\pi}\int_{\phi(S^2)}\big\{A\beta_{\bar{i}}\phi_{\bar{j}}d\bar{\phi}^{\bar{j}}\wedge d\bar{\phi}^{\bar{i}}+[B(\beta\bar{\phi})\delta_{i\bar{j}}+C(\beta\bar{\phi})\bar{\phi}_i\phi_{\bar{j}}
+D\bar{\phi}_i\beta_{\bar{j}}]d\bar{\phi}^{\bar{j}}\wedge d\phi^i\big\}\,,\nonumber
\end{eqnarray}
where the functions $A,\ B,\ C$ and $D$ are
\begin{eqnarray}
A(r^2)&=&-\frac{1}{4}\,\rho r^2F\Big(F^2+2\,\frac{dF}{dr^2}\Big)\,,
\nonumber\\[2mm]
B(r^2)&=&NG+F(1-Fr^2)\,,
\nonumber\\[2mm]
C(r^2)&=&N\,\frac{dG}{dr^2}+\Big(2\,\frac{dF}{dr^2}-F^2\Big)\Big(1-\frac{1}{2}\,\rho F\Big)+\frac{F}{r^2}\Big(1-2\rho F-2r^4\frac{dF}{dr^2}\Big)\,,\nonumber\\[2mm]
D(r^2)&=&\frac{1}{2}\,\rho r^2F\Big(2\frac{dF}{dr^2}-F^2\Big)+2\rho F^2\,.
\end{eqnarray}
One can simplify Eq.\,(\ref{CPNLanomaly}) integrating by parts. First note
\begin{eqnarray}
&&\int_{\phi(S^2)}A(r^2)\beta_{\bar{i}}\phi_{\bar{j}}d\bar{\phi}^{\bar{j}}\wedge d\bar{\phi}^{\bar{i}}\nonumber\\[2mm]
&&=\int_{\phi(S^2)}A(r^2)\beta_{\bar{i}}dr^2\wedge d\bar{\phi}^{\bar{i}}-A(r^2)\beta_{\bar{i}}\bar{\phi}_{j}d\phi^{j}\wedge d\bar{\phi}^{\bar{i}}\nonumber\\[2mm]
&&=\int_{\phi(S^2)}d\, \Big(\, \frac{1}{\rho}+\log\rho-2\log(1+\rho)\beta_{\bar{i}}d\bar{\phi}^{\bar{i}}\,\Big)
+A(r^2)\beta_{\bar{j}}\bar{\phi}_id\bar{\phi}^{\bar{j}}\wedge d\phi^i\nonumber\\[2mm]
&&=\int_{\phi(S^2)}A(r^2)\beta_{\bar{j}}\bar{\phi}_id\bar{\phi}^{\bar{j}}\wedge d\phi^i\,.
\end{eqnarray}
In addition, for the function $C$ term,
\begin{eqnarray}
&&\int_{\phi(S^2)}C(r^2)(\beta\bar{\phi})\bar{\phi}_i\phi_{\bar{j}}d\bar{\phi}^{\bar{j}}\wedge d\phi^i\nonumber\\[2mm]
&&=\int_{\phi(S^2)}C(r^2)(\beta\bar{\phi})\bar{\phi}_idr^2\wedge d\phi^i\nonumber\\[2mm]
&&=\int_{\phi(S^2)}d\frac{Nr^2+2(\rho-1)}{2\rho^2 r^2}(\beta\bar{\phi})\bar{\phi}_id\phi^i\nonumber\\[2mm]
&&=\int_{\phi(S^2)}\frac{-Nr^2-2(\rho-1)}{2\rho^2 r^2}[(\beta\bar{\phi})\delta_{i\bar{j}}+\beta_{\bar{j}}\bar{\phi}_i]d\bar{\phi}^{\bar{j}}\wedge d\phi^i\nonumber\\[2mm]
&&\equiv\int_{\phi(S^2)}\tilde{C}(r^2)[(\beta\bar{\phi})\delta_{i\bar{j}}+\beta_{\bar{j}}\bar{\phi}_i]d\bar{\phi}^{\bar{j}}\wedge d\phi^i\,.
\end{eqnarray}
Combining the above two terms into Eq.(\ref{CPNLanomaly}), one can find
\begin{eqnarray}
\delta_{\beta}\Gamma_{eff}&=&-\frac{1}{4\pi}\int_{\phi(S^2)}\big\{[B-\tilde{C}](\beta\bar{\phi})\delta_{i\bar{j}}+[A+D
-\tilde{C}]\beta_{\bar{j}}\bar{\phi}_i\big\}d\bar{\phi}^{\bar{j}}\wedge d\phi^i\nonumber\\[2mm]
&=&\frac{N}{4\pi}\int_{\phi(S^2)}\frac{\beta_{\bar{j}}\bar{\phi}_i}{2(1+\bar{\phi}\phi)}\,d\bar{\phi}^{\bar{j}}\wedge d\phi^i\nonumber\\[2mm]
&=&-\frac{i}{8\pi}\int_{\phi(S^2)}(\beta\bar{\phi})\,c_1\,.
\label{CPNLanomaly2}
\end{eqnarray}
We also need to add the variation of action with respect to $\bar{\beta}$, which is obtained by Hermitian conjugation of Eq.(\ref{CPNLanomaly2}). Finally we have the result identical  with Eq.(\ref{CPGLanomaly}),
\beq
\mathcal{I}_\beta=\frac{i}{8\pi}\int(\bar{\beta}\phi-\beta\bar{\phi})c_1\,.
\eeq

\end{document}